\documentclass[a4paper]{article}
\usepackage[utf8]{inputenc}

\title{%
  \texorpdfstring{%
    Implementing CNN Layers on the Manticore Cluster-Based Many-Core Architecture \\
    \vspace*{.5\baselineskip}\large%
    Technical Report
  }{%
    Implementing CNN Layers on the Manticore Cluster-Based Many-Core Architecture
  }
}
\author{%
  \texorpdfstring{%
    Andreas Kurth, Fabian Schuiki, and Luca Benini\\
    \small%
    Digital Circuits and Systems Group, ETH Zurich
  }{%
    Andreas Kurth, Fabian Schuiki, and Luca Benini
  }
}
\date{v1.0, April 2021}

\usepackage{algorithm2e}
\usepackage{amsfonts}
\usepackage[shortlabels]{enumitem}
\usepackage[acronym]{glossaries}
\usepackage{IEEEtrantools}
\usepackage{mathtools}
\usepackage{microtype}
\usepackage[binary-units,per-mode=symbol,detect-all]{siunitx}
\usepackage{xcolor}
\usepackage[hidelinks,pdfusetitle]{hyperref} %
\usepackage[capitalize]{cleveref} %

\DeclarePairedDelimiter{\ceil}{\lceil}{\rceil}
\newcommand*{\pluseq}{\ensuremath{\mathrel{+}=}\xspace}

\newcommand*{\BatchSize}{\ensuremath{B}\xspace}
\newcommand*{\ClusterId}{\ensuremath{\mathit{CID}}\xspace}
\newcommand*{\ClusterIdLTwoQuad}{\ensuremath{\ClusterId\sb{\mathrm{in\,L2\,quad}}}\xspace}
\newcommand*{\FiltParams}{\ensuremath{\mathbf{F}}\xspace}
\newcommand*{\FiltStride}{\ensuremath{S}\xspace}
\newcommand*{\FiltWidth}{\ensuremath{F}\xspace}
\newcommand*{\FiltParamsFullConv}{\ensuremath{%
    \FiltParams[\FiltWidth \times \FiltWidth \times \InpDepth \times \OupDepth]%
  }\xspace}
  \newcommand*{\FiltParamsFullFc}{\ensuremath{%
    \FiltParams[\InpWidth \times \InpWidth \times \InpDepth \times \OupDepth]%
  }\xspace}
\newcommand*{\InpDepth}{\ensuremath{D\sb{I}}\xspace}
\newcommand*{\InpVol}{\ensuremath{\mathbf{I}}\xspace}
\newcommand*{\InpVolFull}{\ensuremath{%
    \InpVol[\InpWidth \times \InpWidth \times \InpDepth]%
  }\xspace}
\newcommand*{\InpWidth}{\ensuremath{W\sb{I}}\xspace}
\newcommand*{\OupDepth}{\ensuremath{D\sb{O}}\xspace}
\newcommand*{\OupDepthStack}{\ensuremath{\Delta\sb{O}}\xspace}
\newcommand*{\OupVol}{\ensuremath{\mathbf{O}}\xspace}
\newcommand*{\OupVolFullConv}{\ensuremath{%
    \OupVol[\OupWidth \times \OupWidth \times \OupDepth]%
  }\xspace}
\newcommand*{\OupVolFullFc}{\ensuremath{%
    \OupVol[1 \times 1 \times \OupDepth]%
  }\xspace}
\newcommand*{\OupWidth}{\ensuremath{W\sb{O}}\xspace}
\newcommand*{\ZeroPadding}{\ensuremath{P}\xspace}

\SetKwFor{ParFor}{parallelize for}{over clusters do}{end}
\SetKwFunction{Conv}{Conv}
\SetKwFunction{DmaLoad}{DmaLoad}
\SetKwFunction{DmaStore}{DmaStore}
\SetKwFunction{DmaWait}{DmaWait}
\SetKwFunction{ElemMac}{ElemMac}
\newcommand{\AlgFooter}[1]{%
  \BlankLine
  \caption{#1}%
}
\newcommand{\AlgHeaderConv}{%
  \KwIn{\InpVolFull input volume}
  \KwIn{\FiltParamsFullConv filter parameters}
  \KwOut{\OupVolFullConv output volume}
  \BlankLine
}
\newcommand{\AlgHeaderFc}{%
  \KwIn{\InpVolFull input volume}
  \KwIn{\FiltParamsFullFc filter parameters}
  \KwOut{\OupVolFullFc output volume}
  \BlankLine
}

\newacronym{ccr}{CCR}{compute-to-communication ratio}
\newacronym{cnn}{CNN}{convolutional neural network}
\newacronym{dma}{DMA}{direct memory access}
\newacronym{fpu}{FPU}{floating-point unit}
\newacronym{mac}{MAC}{multiply-and-accumulate}
\newacronym{rte}{RTE}{runtime environment}
\newacronym[longplural={scratchpad memories}]{spm}{SPM}{scratchpad memory}

\DeclareSIUnit{\mac}{MAC}
\DeclareSIUnit{\word}{word}
\DeclareSIUnit{\spflop}{spflop}
\DeclareSIUnit{\dpflop}{dpflop}

\newcommand{\BufStorage}[3]{%
  Beyond the minimum local memory requirements, additional storage can be required to hide the latency of loads from off-chip main memory.
  Especially #1, which are loaded once per #2, can benefit from a buffer.
  The size of that buffer depends on the latency to the main memory.
  Assuming a round-trip latency of 256 clock cycles and a data bus width of $ \SI{512}{\bit} = \SI{64}{\byte} $, up to \SI{16}{\kibi\byte} can be in transfer in the on-chip network.
  A buffer of the same size can be added for #3, so that data transfers by the \gls{dma} engine run fully in background.\footnote{%
    The \acrshort{dma} transfer buffer could be shared by the input depth slices and the filter parameters, since the total amount of data in transfer in the on-chip network does not depend on the type of data being transferred.
    This would save local memory in the cluster, but it requires the \acrshort{rte} to dynamically partition the \acrshort{dma} transfer buffer between different data types and variables, which is not trivial.
    We therefore do not assume this \acrshort{rte} capability.
  }
}

\begin{document}

\maketitle

\section*{Abstract}
This document presents implementations of fundamental \gls{cnn} layers on the Manticore cluster-based many-core architecture and discusses their characteristics and trade-offs.

\glsresetall

\section{Introduction}

\begin{figure}
  \centering
  \includegraphics[width=\textwidth]{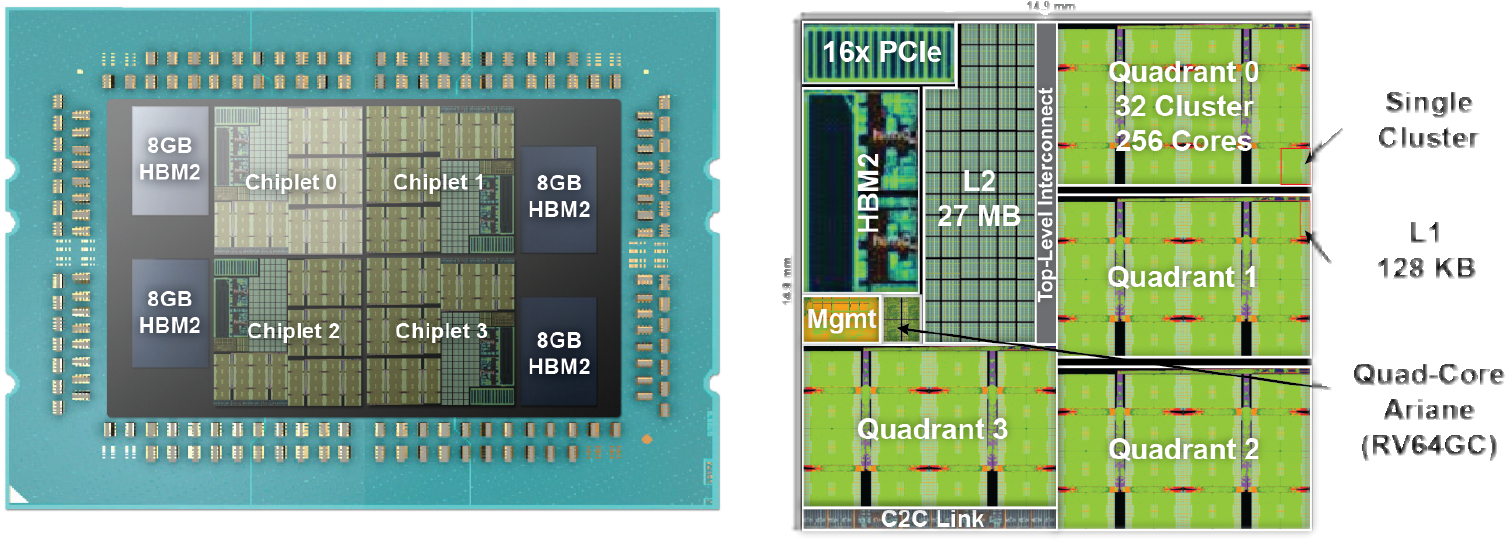}
  \caption{Conceptual floorplan of Manticore's package (left) and an individual chiplet (right)~\cite{manticore}.}%
  \label{fig:manticore_chiplet}
\end{figure}

The \emph{Manticore} architecture~\cite{manticore} is a many-core processor designed for high-performance, high-efficiency data-parallel floating-point computing.
The implementation proposed in \cite{manticore} consists of four chiplets on an interposer, see \cref{fig:manticore_chiplet}.
Each chiplet contains 1024 cores grouped into 128 clusters, one \SI{8}{\gibi\byte} HBM2E controller and PHY, as well as L2 memory and I/Os to connect to the other chiplets.

\begin{figure}
  \centering
  \includegraphics[width=.6\textwidth]{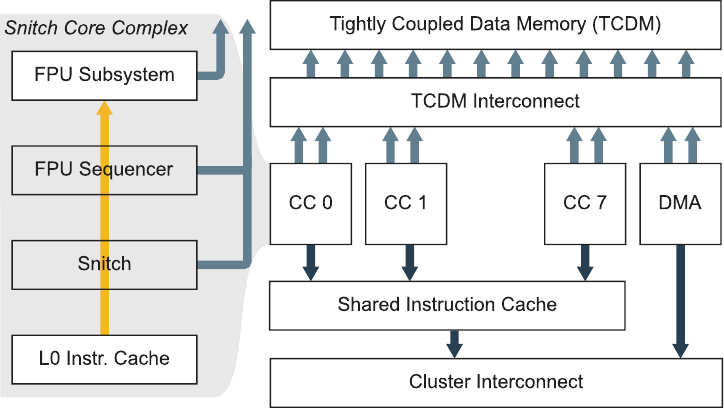}
  \caption{Simplified block diagram of a Manticore cluster~\cite{manticore}.}%
  \label{fig:manticore_cluster}
\end{figure}

Each cluster, shown in \cref{fig:manticore_cluster}, contains eight small 32-bit integer RISC-V cores~\cite{zaruba2020snitch}, each paired with a large double-precision \gls{fpu}~\cite{mach2019fpu}, and \SI{128}{\kibi\byte} of tightly-coupled L1 \gls{spm} organized in 32 banks of \SI{64}{\bit} width each.
The \gls{fpu} is capable of computing one double-precision floating-point \gls{mac} operation or two single-precision \gls{mac} operations in each clock cycle.
As primary means for moving data into and out of L1, each cluster contains two 512-bit-wide \gls{dma} engines~\cite[Section~2.5]{kurth2020axi} -- one for reads and one for writes -- which are attached to the L1 memory and share one master port into the on-chip network.
\Gls{dma} engines in other clusters can access the L1 memory through an additional 512-bit-wide slave port.
Each cluster has a 64-bit master port to let its cores access external memory and a 64-bit slave port to let cores in other clusters access its L1 memory.
Four clusters form an L1 quadrant, four L1 quadrants form an L2 quadrant, four L2 quadrants form an L3 quadrant, and two L3 quadrants form a chiplet.

\begin{figure}
  \centering
  \includegraphics[width=.8\textwidth]{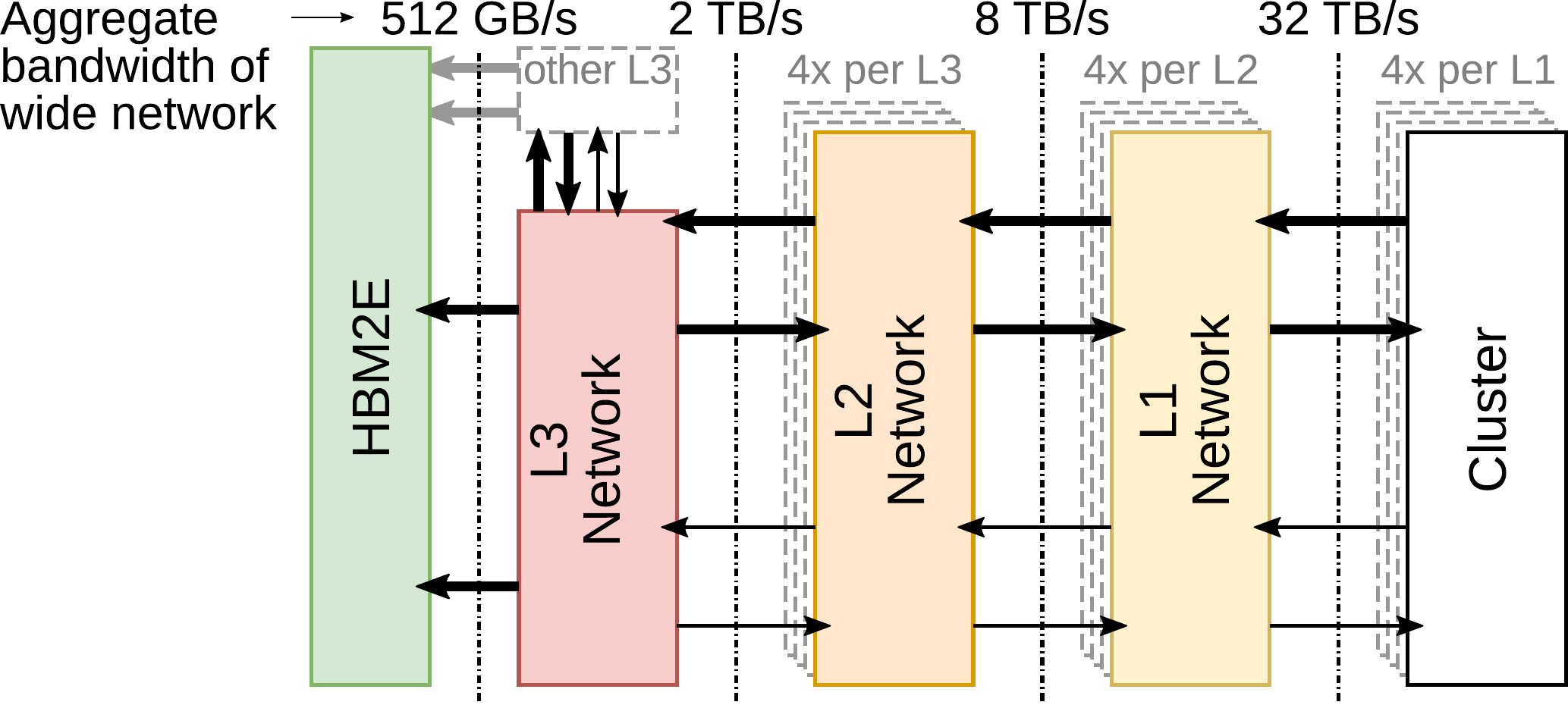}
  \caption{On-chip network of one Manticore chiplet, adapted from \cite[Section~4]{kurth2020axi}.}%
  \label{fig:manticore_noc}
\end{figure}

Manticore's on-chip network~\cite[Section~4]{kurth2020axi}, shown in \cref{fig:manticore_noc}, features:
\begin{enumerate}[(1)]
  \item physically separate networks for traffic by \gls{dma} engines and cores to minimize the interference between the wide bursts of \gls{dma} engines and the word-wise accesses of the cores;
  \item a tree topology to combine a high bandwidth between off-chip main memory and any cluster for effective data transfers and low latency between any two cores for efficient concurrency;
  \item fully-connected crossbars within each quadrant to provide units within the same quadrant with a high bandwidth for effective local data sharing; and
  \item the same data width and frequency throughout the \gls{dma} network to provide a high bandwidth between off-chip main memory and any clusters.
\end{enumerate}

This document describes the implementation of convolutional and fully-connected layers, which together account for \SIrange{95}{99}{\percent} of the floating-point operations in \glspl{cnn}, on a Manticore chiplet.

\subsection{Convolutional Neural Networks (CNNs)}

We follow the notation used in Stanford University's ``CS231n Convolutional Neural Networks for Visual Recognition''~\cite[Module~2: CNNs]{cs231n}.
There is no single established convention on describing neural networks, but adapting the notation of a course rather than that of a specific paper in the field hopefully makes this document understandable for a wide range of readers.

\begin{figure}
  \centering
  \includegraphics[width=.6\textwidth]{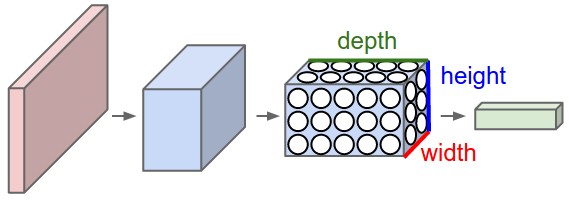}
  \caption{%
    A \acrfull{cnn} with its input layer in red on the left, two hidden layers in blue in the middle, and its output layer in green on the right~\cite{cs231n}.
    Every layer of a \acrshort{cnn} transforms a 3D input volume of neuron activations to a 3D output volume of neuron activations.
    The neurons of each layer are arranged in three dimensions: width, height, and depth.
  }%
  \label{fig:cnn}
\end{figure}

A \gls{cnn} is generally a sequence of layers, and each \emph{layer} transforms one three-dimensional \emph{volume} of activations to another three-dimensional volume of activations.
The layers of a \gls{cnn}, shown in \cref{fig:cnn}, have neurons arranged in three dimensions: width, height, and depth.
\emph{Width} and \emph{height} are the \emph{spatial dimensions}, and \emph{depth} is the third dimension of a layer (and not the number of layers in a network).

The neurons in convolutional and fully-connected layers are also called \emph{filter parameters}, because they are the parameters of mathematical filters that are applied to transform an input volume to an output volume.

The \emph{hyperparameters} of a layer, listed in \cref{tbl:cnn_hyperparameters}, describe the shape of its input and output volumes and how its filter is applied.

\begin{table}
  \centering
  \begin{tabular}{ l l }
    \InpWidth & width and height of input volume \\
    \InpDepth & depth of input volume \\
    \OupWidth & width and height of output volume \\
    \OupDepth & depth of output volume \\
    \FiltWidth & receptive field (i.e., width and height) of the filter \\
    \ZeroPadding & zero padding elements in the spatial dimensions \\
    \FiltStride & stride of the filter \\
    \BatchSize & batch size \\
  \end{tabular}
  \caption{Hyperparameters of \acrshort{cnn} layers.}%
  \label{tbl:cnn_hyperparameters}
\end{table}

A convolutional layer transforms an input volume \InpVolFull (i.e., of width and height \InpWidth and depth \InpDepth) to an output volume \OupVolFullConv.
It does this transformation by convolving every two-dimensional \emph{depth slice} (i.e., a matrix with \InpWidth rows and columns) of the input volume with a two dimensional filter kernel (i.e., a matrix with \FiltWidth rows and columns) to produce one two-dimensional depth slice (i.e., a matrix with \OupWidth rows and columns) of the output volume.
Each pair of input and output depth slice has its own filter kernel.
The overall filter parameters of one layer (also known as the neurons of the convolutional layer) are thus \FiltParamsFullConv.
Before the convolution with the filter kernel, each input depth slice is \emph{padded} with \ZeroPadding zeros at the beginning and end of each row and column.
The convolutional filter is then applied with a \emph{stride} \FiltStride, i.e., to every \FiltStride{}-th row and column.
The width and height of the output volume are thus given by $\OupWidth = (\InpWidth + 2 \ZeroPadding - \FiltWidth) / \FiltStride + 1$.

A fully-connected layer transforms an input volume \InpVolFull to an output volume \OupVolFullFc.
Any fully-connected layer can be represented as convolutional layer by setting $\FiltWidth = \InpWidth$ (and $\FiltStride = 1$, $\ZeroPadding = 0$), so the filter parameters of a convolutional layer are \FiltParamsFullFc.

\subsection{Notation}

\subsubsection{Algorithms}

\textit{For} loops are written as ``\textbf{for} $\mathit{var} \leftarrow \mathit{lb}$ \KwTo $\mathit{ub}$ \textbf{do} $\mathit{body}$ \textbf{end}'', where the loop variable $\mathit{var}$ is initialized to $\mathit{lb}$ and is incremented in each iteration by one.
The loop $\mathit{body}$ is executed as long as $\mathit{var} < \mathit{ub}$.

\textit{For} loops executed in parallel by all clusters start with ``\textbf{parallelize for}'' and their loop header ends with ``\textbf{over clusters do}''.
The schedule of the parallelization is not restricted and may be optimized by the compiler, the \gls{rte}, or both.
In this document, we do not detail parallel execution \emph{within} clusters.

Indices into multi-dimensional volumes are written in square brackets with dimensions separated by commas (e.g., $[a, b, c]$), where the innermost dimension is on the left and the outermost dimension is on the right.
Indexing starts at zero.
A range of indexes is written with a colon, e.g., $[d:e]$ is a one-dimensional range including all elements starting at $d$ and ending before $e$, i.e., $\{n \in \mathbb{N} \mid d \leq n < e\}$.
A range without start and end, such as $[:]$, includes all elements in a dimension.

Data is loaded from main memory into cluster local memory and stored back primarily through \gls{dma} engine transfers.
Such loads and stores are denoted with \DmaLoad{} and \DmaStore{}, respectively.
\Gls{dma} engine transfers run asynchronous to computation.
That is, the cores of the cluster can continue to compute while the \gls{dma} engine transfers data.
Of course, the cores can only compute on data in local memory that has already been transferred.
To ensure synchronization between cores and \gls{dma} engine where necessary, the \DmaWait{} function blocks until the data given in the argument is completely transferred.

Each cluster has a constant, unique numerical ID.
Numbering starts at $0$ for the first cluster in the first L1 quadrant of the first L2 quadrant of the first L3 quadrant.
The next cluster cluster in that same L1 quadrant has ID $1$, the first cluster in the next L1 quadrant of the same L2 quadrant has ID $4$, and so on.
The cluster ID is denoted as \ClusterId.

\subsubsection{Units}

Single- and double-precision floating-point numbers and operations according to IEEE~754 are both supported by Manticore.
In this document, we assume that all numbers in a layer have either single- or double-precision.
This allows for a generic discussion where each number is a \emph{word}.
A single-precision word is \SI{4}{\byte} (four byte) in size, and a double-precision word is \SI{8}{\byte} (eight byte) in size.

The main compute operation in \gls{cnn} layers are \emph{\glsreset{mac}\glspl{mac}}.
We count one \gls{mac} as two floating-point operations (since it consists of one multiplication and one addition).
An operation on single-precision numbers is called \emph{single-precision floating-point operation (\si{\spflop})}, and an operation on double-precision numbers is called \emph{double-precision floating-point operation (\si{\dpflop})}.

\section{Convolutional Layers}

\subsection{Parallelize Output Depth Slices over Clusters}

\begin{algorithm}[htbp]
  \AlgHeaderConv

  \ParFor{$d\sb{o} \leftarrow 0$ \KwTo $\OupDepth$}{%
    initialize $\OupVol[:,:,d\sb{o}]$ in local memory to zero\;
    \DmaLoad{$\InpVol[:,:,0]$} from main memory\;
    \DmaLoad{$\FiltParams[:,:,0,d\sb{o)}]$} from main memory\;
    \For{$d\sb{i} \leftarrow 0$ \KwTo $\InpDepth$}{%
      \If{$d\sb{i} + 1 < \InpDepth$}{%
          \DmaLoad{$\InpVol[:,:,d\sb{i}+1]$} from main memory\;
          \DmaLoad{$\FiltParams[:,:,d\sb{i}+1,d\sb{o)}]$} from main memory\;
      }
      \DmaWait{$\InpVol[:,:,d\sb{i}]$}\;
      \DmaWait{$\FiltParams[:,:,d\sb{i},d\sb{o)}]$}\;
      $\OupVol[:,:,d\sb{o}]$ \pluseq \Conv{$\InpVol[:,:,d\sb{i}]$, $\FiltParams[:,:,d\sb{i},d\sb{o}]$}\;
    }
    \DmaStore{$\OupVol[:,:,d\sb{o}]$} to main memory\;
  }

  \AlgFooter{Implementation of a convolutional layer where the depth slices of the output volumes are parallelized over the clusters.}%
  \label{alg:conv:par_oup_depth_slices}
\end{algorithm}

The implementation of a convolutional layer shown in \cref{alg:conv:par_oup_depth_slices} parallelizes the output depth slices over the clusters.
In the parallel section, the cluster first initializes the output depth slice in its local memory to zero and then loops over the input depth slices.
Within that loop, the cluster loads the input depth slice and the filter parameters for the current pair of input and output depth slices from main memory.\footnote{%
  The loading of input data is split into two parts:
  Before the loop over the input depth slices, the data for the first iteration of the loop is loaded.
  Within the loop, the data for the subsequent iteration of the loop is loaded (except for the last iteration).
  The calls to \DmaWait{} make sure that the data for the current iteration is present in local memory before the computation. 
}
Then the cluster convolves the input depth slice with the filter parameters and accumulates the result to the output depth slice in local memory.
After it has looped over all input depth slices, the cluster stores the output depth slice to main memory.

This implementation exploits the independence of the output depth slices, necessitating no communication between the clusters.

The main drawback of this implementation is that each cluster loads all input depth slices for each output depth slice it processes, because it cannot store the entire input volume in local memory.
This leads to a lot of data loaded over and over again, which significantly reduces the \gls{ccr} and causes the implementation to become memory-bound.
The implementation in \cref{sec:conv:par_oup_depth_slices_stack} will resolve this limitation by processing output depth slices in stacks.

\subsubsection{Compute Complexity}
The convolution (\Conv{}) does $\InpWidth \cdot \InpWidth \cdot \FiltWidth\sp{2}$ \glspl{mac}, and it is placed inside a loop with \InpDepth iterations.
In total, each cluster task does $\InpWidth\sp{2} \cdot \FiltWidth\sp{2} \cdot \InpDepth$ \glspl{mac}.
Together, the clusters do $\InpWidth\sp{2} \cdot \FiltWidth\sp{2} \cdot \InpDepth \cdot \OupDepth$ \glspl{mac}.

\subsubsection{Space Complexity}%
\label{sec:conv:par_oup_depth_slices:space}
Each cluster must store:
\begin{enumerate}[(1)]
  \item its output depth slice, which are $\OupWidth\sp{2}$ words;
  \item one input depth slice, which are $\InpWidth\sp{2}$ words;
  \item the filter parameters for one input-output depth slice pair, which are $\FiltWidth\sp{2}$ words.
\end{enumerate}
Thus, the minimum local memory required per cluster is $\OupWidth\sp{2} + \InpWidth\sp{2} + \FiltWidth\sp{2}$ words.
For a typical layer with $\InpWidth = 32$, $\FiltWidth = 3$, and $\OupWidth = 32$ (due to $\FiltStride = 1$ and $\ZeroPadding = 1$), this corresponds to \num{2057} words, or less than \SI{8.1}{\kibi\byte} for single-precision and \SI{16.1}{\kibi\byte} for double-precision data.
Both easily fit into the \SI{128}{\kibi\byte} of local memory of each cluster.

\BufStorage%
  {the input depth slice (2) and the filter parameters (3)}%
  {inner loop iteration}%
  {output depth slice (1)}
In total, roughly \SI{48}{\kibi\byte} are required as buffers.

\subsubsection{Communication Complexity}%
\label{sec:conv:par_oup_depth_slices:communication}

\paragraph{Inter-Cluster Communication}
No communication between the clusters is required.
The clusters exclusively access data that is not shared among the clusters in main memory.

\paragraph{Main Memory Accesses}
Inside the parallel section, each cluster executes a loop of \InpDepth iterations.
Inside that loop, each cluster loads one input depth slice, which are $\InpWidth\sp{2}$ words, and the filter parameters for one input-output depth slice pair, which are $\FiltWidth\sp{2}$ words, from main memory.
Over all loop iterations, each cluster thus loads $\InpDepth \cdot (\InpWidth\sp{2} + \FiltWidth\sp{2})$ words from main memory.
After the loop, each cluster stores one output depth slice, which are $\OupWidth\sp{2}$ words, to main memory.
Together, the clusters in all iterations load $\OupDepth \cdot \InpDepth \cdot (\InpWidth\sp{2} + \FiltWidth\sp{2})$ words from and store $\OupDepth \cdot \OupWidth\sp{2}$ words to main memory.

\subsubsection{Compute-to-Communication Ratio}%
\label{sec:conv:par_oup_depth_slices:ccr}

Dividing the number of \glspl{mac} by the number of words accessed in main memory gives the \gls{ccr}:
\begin{IEEEeqnarray}{rCl}
  \text{\acrshort{ccr}} & = & \frac{\InpWidth\sp{2} \cdot \FiltWidth\sp{2} \cdot \InpDepth \cdot \OupDepth}{\OupDepth \cdot \InpDepth \cdot (\InpWidth\sp{2} + \FiltWidth\sp{2}) + \OupDepth \cdot \OupWidth\sp{2}} \frac{\si{\mac}}{\si{\word}} \\
  & = & \frac{\InpDepth \cdot \InpWidth\sp{2} \cdot \FiltWidth\sp{2}}{\InpDepth \cdot (\InpWidth\sp{2} + \FiltWidth\sp{2}) + \OupWidth\sp{2}} \frac{\si{\mac}}{\si{\word}}.
\end{IEEEeqnarray}

To give some numerical intuition:
For typical values $\InpWidth = \OupWidth = 32$, $\InpDepth = \OupDepth = 128$, and $\FiltWidth = 3$, the \gls{ccr} is ca.\ \SI{8.9}{\mac\per\word}.
That is, the \gls{ccr} for single-precision data is \SI{4.4}{\spflop\per\byte} and for double-precision data \SI{2.2}{\dpflop\per\byte}.
These are very low values, which cause \cref{alg:conv:par_oup_depth_slices} to be memory-bound on Manticore.

The reason is that the algorithm reloads each input depth slice once per output depth slice.
Thus, the \gls{ccr} is mainly determined by the receptive field of the filter, \FiltWidth.
This can also be seen analytically as follows:
Make the reasonable assumption that $\OupWidth = \InpWidth$ (i.e., stride \FiltStride and padding \ZeroPadding are set accordingly).
Then the \gls{ccr} becomes
\begin{IEEEeqnarray}{rCl}
  \text{\acrshort{ccr}} & \stackrel{\OupWidth = \InpWidth}{=} & \frac{\InpDepth \cdot \InpWidth\sp{2} \cdot \FiltWidth\sp{2}}{\InpDepth \cdot (\InpWidth\sp{2} + \FiltWidth\sp{2}) + \InpWidth\sp{2}} \frac{\si{\mac}}{\si{\word}} \\
  & = & \frac{\InpWidth\sp{2} \cdot \InpDepth \cdot \FiltWidth\sp{2}}{\InpWidth\sp{2} \cdot (\InpDepth + 1) + \FiltWidth\sp{2} \cdot \InpDepth} \frac{\si{\mac}}{\si{\word}}.
\end{IEEEeqnarray}
Typically, $\InpDepth \gg 1$, so
\begin{equation}
  \text{\acrshort{ccr}} \stackrel{\InpDepth \gg 1}{\approx} \frac{\InpDepth \cdot \InpWidth\sp{2} \cdot \FiltWidth\sp{2}}{\InpDepth \cdot (\InpWidth\sp{2} + \FiltWidth\sp{2})} \frac{\si{\mac}}{\si{\word}} = \frac{\InpWidth\sp{2} \cdot \FiltWidth\sp{2}}{\InpWidth\sp{2} + \FiltWidth\sp{2}} \frac{\si{\mac}}{\si{\word}}.
\end{equation}
Finally, typically $\InpWidth\sp{2} \gg \FiltWidth\sp{2}$, thus
\begin{equation}
  \text{\acrshort{ccr}} \stackrel{\InpWidth\sp{2} \gg \FiltWidth\sp{2}}{\approx} \FiltWidth\sp{2} \frac{\si{\mac}}{\si{\word}}.
\end{equation}

\subsubsection{Summary}
\Cref{alg:conv:par_oup_depth_slices} implements convolutional layers in parallel on all clusters of Manticore without necessitating communication or synchronization between the clusters.
The modest memory requirements of this implementation allow to process large layers.
The main drawback of this implementation is that each cluster loads the entire input volume once for each output depth slice it processes.
This leads to a lot of redundant data loads, which significantly reduces the \gls{ccr} and causes this implementation to be memory-bound.

\subsection{Parallelize Stacks of Output Depth Slices over Clusters}%
\label{sec:conv:par_oup_depth_slices_stack}

\begin{algorithm}[htbp]
  \AlgHeaderConv

  \ParFor{$\delta\sb{o} \leftarrow 0$ \KwTo $\ceil{\OupDepth / \OupDepthStack}$}{%
    $D\sb{O,\mathrm{begin}} \leftarrow \delta\sb{o} \cdot \OupDepthStack$\;
    $D\sb{O,\mathrm{end}} \leftarrow \min(D\sb{O,\mathrm{begin}} + \OupDepthStack, \OupDepth) $\;
    initialize $\OupVol[:,:,D\sb{O,\mathrm{begin}}:D\sb{O,\mathrm{end}}]$ in local memory to zero\;
    \DmaLoad{$\InpVol[:,:,0]$} from main memory\;
    \For{$d\sb{i} \leftarrow 0$ \KwTo $\InpDepth$}{%
      \If{$d\sb{i} + 1 < \InpDepth$}{%
        \DmaLoad{$\InpVol[:,:,d\sb{i}+1]$} from main memory\;
      }
      \DmaLoad{$\FiltParams[:,:,d\sb{i},D\sb{O,\mathrm{begin}}]$} from main memory\;
      \DmaWait{$\InpVol[:,:,d\sb{i}]$}\;
      \For{$d\sb{o} \leftarrow D\sb{O,\mathrm{begin}}$ \KwTo $D\sb{O,\mathrm{end}}$}{%
        \If{$d\sb{o} + 1 < D\sb{O,\mathrm{end}}$}{%
          \DmaLoad{$\FiltParams[:,:,d\sb{i},d\sb{o}+1]$} from main memory\;
        }
        \DmaWait{$\FiltParams[:,:,d\sb{i},d\sb{o}]$}\;
        $\OupVol[:,:,d\sb{o}]$ \pluseq \Conv{$\InpVol[:,:,d\sb{i}]$, $\FiltParams[:,:,d\sb{i},d\sb{o}]$}\;
      }
    }
    \DmaStore{$\OupVol[:,:,D\sb{O,\mathrm{begin}}:D\sb{O,\mathrm{end}}]$} to main memory\;
  }

  \AlgFooter{Implementation of a convolutional layer where stacks of depth slices of the output volumes are parallelized over the clusters.}%
  \label{alg:conv:par_oup_depth_slices_stack}
\end{algorithm}

The implementation of a convolutional layer shown in \cref{alg:conv:par_oup_depth_slices_stack} extends \cref{alg:conv:par_oup_depth_slices} by processing output depth slices in stacks:  each cluster computes \OupDepthStack output depth slices in each parallel task.
Within the parallel task, the stack of output depth slices is initialized in the beginning and stored to main memory at the end as a whole.
Between that, output depth slices are calculated as in \cref{alg:conv:par_oup_depth_slices}.
This stacking of output depth slices increases the reuse factor of each input depth slice and therefore reduces the \gls{ccr} compared to \cref{alg:conv:par_oup_depth_slices}.

\subsubsection{Compute Complexity}%
\label{sec:conv:par_oup_depth_slices_stack:compute}
The convolution (\Conv{}) does $\InpWidth\sp{2} \cdot \FiltWidth\sp{2}$ \glspl{mac}, and it is placed inside an inner loop with \OupDepthStack iterations.
(If \OupDepthStack does not evenly divide \OupDepth, one cluster task processes $\OupDepth \bmod \OupDepthStack$ instead of \OupDepthStack depth slices.)
The inner loop is placed inside an outer loop with \InpDepth iterations.
Thus, one cluster task in the common case does $\InpWidth\sp{2} \cdot \FiltWidth\sp{2} \cdot \OupDepthStack \cdot \InpDepth$ \glspl{mac}, and all clusters together do
$\InpWidth\sp{2} \cdot \FiltWidth\sp{2} \cdot \OupDepth \cdot \InpDepth$ \glspl{mac}.
Thus, in comparison to \cref{alg:conv:par_oup_depth_slices}, this algorithm does not add additional \glspl{mac}.

\subsubsection{Space Complexity}%
\label{sec:conv:par_oup_depth_slices_stack:space}
Each cluster must store as many input depth slices and filter parameters as described in \cref{sec:conv:par_oup_depth_slices:space}, but instead of one output depth slice, it must now store \OupDepthStack output depth slices.
Thus, the minimum local memory required per cluster is $\OupDepthStack \cdot \OupWidth\sp{2} + \InpWidth\sp{2} + \FiltWidth\sp{2}$ words.

With \SI{32}{\kibi\byte} of local memory reserved for buffers for the input depth slice and the filter parameters (see \cref{sec:conv:par_oup_depth_slices:space}), \SI{96}{\kibi\byte} are left for the stack of output depth slices.
To give some numerical intuition, for a typical layer with $\OupWidth = 32$, this limits $\OupDepthStack \leq 24$ for single-precision data and $\OupDepthStack \leq 12$ for double-precision data.

\subsubsection{Communication Complexity}

\paragraph{Inter-Cluster Communication}
As for \cref{alg:conv:par_oup_depth_slices}, no communication between the clusters is required.

\paragraph{Main Memory Accesses}
The accesses to main memoryBy cleverly using the local memory of each cluster and the high-performance on-chip network, two convolutional layer implementations and one fully-connected layer implementation attain a high \gls{ccr}, w

 are similar to those discussed in \cref{sec:conv:par_oup_depth_slices:communication}, with one important difference:
Whereas \cref{alg:conv:par_oup_depth_slices} loaded each input depth slice once \emph{per output depth slice}, \cref{alg:conv:par_oup_depth_slices_stack} loads each input depth slice once \emph{per stack of output depth slices}.
The total number of words loaded from main memory is thus:
\begin{equation}
  \underbrace{\ceil*{\frac{\OupDepth}{\OupDepthStack}} \cdot \InpDepth \cdot \InpWidth\sp{2}}\sb{\text{input depth slices}} + \underbrace{\OupDepth \cdot \InpDepth \cdot \FiltWidth\sp{2}}\sb{\text{filter parameters}}.
\end{equation}
The total number of words stored to main memory remains $\OupDepth \cdot \OupWidth\sp{2}$.

\subsubsection{Compute-to-Communication Ratio}
Dividing the number of \glspl{mac} by the number of words accessed in main memory gives the \gls{ccr}:
\begin{equation}
  \acrshort{ccr} = \frac{\InpWidth\sp{2} \cdot \FiltWidth\sp{2} \cdot \InpDepth \cdot \OupDepth}{\ceil*{\frac{\OupDepth}{\OupDepthStack}} \cdot \InpDepth \cdot \InpWidth\sp{2} + \OupDepth \cdot \InpDepth \cdot \FiltWidth\sp{2} + \OupDepth \cdot \OupWidth\sp{2}} \frac{\si{\mac}}{\si{\word}}
\end{equation}

To give some numerical intuition for a typical layer with $\InpWidth = 32$, $\FiltWidth = 3$, $\OupWidth = 32$ (due to $\FiltStride = 1$ and $\ZeroPadding = 1$), and $\InpDepth = \OupDepth = 128$:
For single-precision data, \OupDepthStack may be at most 24, and the \gls{ccr} would then be \SI{141.8}{\mac\per\word} or \SI{70.9}{\spflop\per\byte}.
For double-precision data, \OupDepthStack may be at most 12, and the \gls{ccr} would then be \SI{87.8}{\mac\per\word} or \SI{21.9}{\dpflop\per\byte}.
Compared to \cref{alg:conv:par_oup_depth_slices}, reusing each input depth slice for a stack of output depth slices significantly increases the \gls{ccr} and makes this algorithm compute-bound.

\subsubsection{Summary}
\Cref{alg:conv:par_oup_depth_slices_stack} significantly increases the \gls{ccr} compared to \cref{alg:conv:par_oup_depth_slices} by using more of the local memory in each cluster to store and work on a stack of output depth slices at a time.
\Cref{alg:conv:par_oup_depth_slices_stack} maintains the design goal of \cref{alg:conv:par_oup_depth_slices}: still no communication or synchronization is required between the clusters.

\subsection{Parallelize Stacks of Output Depth Slices over Clusters and Reuse Input Depth Slice of Previous Cluster}%
\label{sec:conv:par_oup_depth_slices_stack_inp_depth_slice_reuse}

\begin{algorithm}[htbp]
  \AlgHeaderConv

  \ParFor{$\delta\sb{o} \leftarrow 0$ \KwTo $\ceil{\OupDepth / \OupDepthStack}$}{%
    $D\sb{O,\mathrm{begin}} \leftarrow \delta\sb{o} \cdot \OupDepthStack$\;
    $D\sb{O,\mathrm{end}} \leftarrow \min(D\sb{O,\mathrm{begin}} + \OupDepthStack, \OupDepth) $\;
    initialize $\OupVol[:,:,D\sb{O,\mathrm{begin}}:D\sb{O,\mathrm{end}}]$ in local memory to zero\;
    $\ClusterIdLTwoQuad \leftarrow \ClusterId \bmod 16$\;
    \DmaLoad{$\InpVol[:,:,\ClusterIdLTwoQuad]$} from main memory\;
    \For{$d\sb{i} \leftarrow \ClusterIdLTwoQuad$ \KwTo $\InpDepth$ $\mathrm{then}$ $d\sb{i} \leftarrow 0$ \KwTo \ClusterIdLTwoQuad}{%
      \eIf{$d\sb{i} + 1 = \InpDepth$}{%
        $d\sb{i,\mathrm{next}} \leftarrow 0$\;
      }{%
        $d\sb{i,\mathrm{next}} \leftarrow d\sb{i} + 1$\;
      }
      \If{$d\sb{i,\mathrm{next}} \neq \ClusterIdLTwoQuad$}{%
        \eIf{$d\sb{i,\mathrm{next}} \bmod 16 = \ClusterIdLTwoQuad$}{%
          \DmaLoad{$\InpVol[:,:,d\sb{i,\mathrm{next}}]$} from main memory\;
        }{%
          \DmaLoad{$\InpVol[:,:,d\sb{i,\mathrm{next}}]$} from cluster $(\ClusterId - 1) \bmod 16$\;
        }
      }
      \DmaLoad{$\FiltParams[:,:,d\sb{i},D\sb{O,\mathrm{begin}}]$} from main memory\;
      \DmaWait{$\InpVol[:,:,d\sb{i}]$}\;
      \For{$d\sb{o} \leftarrow D\sb{O,\mathrm{begin}}$ \KwTo $D\sb{O,\mathrm{end}}$}{%
        \If{$d\sb{o} + 1 < D\sb{O,\mathrm{end}}$}{%
          \DmaLoad{$\FiltParams[:,:,d\sb{i},d\sb{o}+1]$} from main memory\;
        }
        \DmaWait{$\FiltParams[:,:,d\sb{i},d\sb{o}]$}\;
        $\OupVol[:,:,d\sb{o}]$ \pluseq \Conv{$\InpVol[:,:,d\sb{i}]$, $\FiltParams[:,:,d\sb{i},d\sb{o}]$}\;
      }
    }
    \DmaStore{$\OupVol[:,:,D\sb{O,\mathrm{begin}}:D\sb{O,\mathrm{end}}]$} to main memory\;
  }

  \AlgFooter{Implementation of a convolutional layer where stacks of depth slices of the output volumes are parallelized over the clusters and the 16 clusters within one L2 quadrant reuse input depth slices.}%
  \label{alg:conv:par_oup_depth_slices_stack_inp_depth_slice_reuse}
\end{algorithm}

The implementation of a convolutional layer shown in \cref{alg:conv:par_oup_depth_slices_stack_inp_depth_slice_reuse} improves \cref{alg:conv:par_oup_depth_slices_stack} in one aspect: instead of loading each input depth slice from main memory, clusters within the same L2 quadrant load every input depth slice that has already been loaded by another cluster in the same L2 quadrant from that cluster.
The main motivation for this is reducing the off-chip traffic to improve the energy efficiency.

\subsubsection{Compute Complexity}
The compute complexity is identical to that of \cref{alg:conv:par_oup_depth_slices_stack}, which is described in \cref{sec:conv:par_oup_depth_slices_stack:compute}.

\subsubsection{Space Complexity}
In addition to the buffers described in \cref{sec:conv:par_oup_depth_slices_stack:space}, this implementation requires a buffer to store the input depth slice so that another cluster can load it.
The round-trip latency within one L2 cluster is in the low tens of cycles, so the dominating factor for this buffer is the size of one input depth slice: $\InpWidth\sp{2}$ words.
This leaves \SI{92}{\kibi\byte} for the stack of output depth slices for single-precision data and \SI{88}{\kibi\byte} for double-precision data.
To give some numerical intuition, for a typical layer with $\OupWidth = 32$, this limits $\OupDepthStack \leq 23$ for single-precision data and $\OupDepthStack \leq 11$ for double-precision data.

\subsubsection{Communication Complexity}
\paragraph{Inter-Cluster Communication}
As there are 16 clusters within one L2 quadrant, in 15 out of 16 iterations over the input depth slices, the input depth slice is loaded from another cluster instead of main memory.
The number of words communicated between clusters is thus:
\begin{equation}
  \frac{15}{16} \cdot \ceil*{\frac{\OupDepth}{\OupDepthStack}} \cdot \InpDepth \cdot \InpWidth\sp{2}.
\end{equation}
To synchronize the update of an input depth slice buffer, the destination cluster sends an atomic increment to the epoch counter in the source cluster.

\paragraph{Main Memory Accesses}
The number of words loaded from main memory is significantly reduced, as input depth slices are mostly loaded from other clusters:
\begin{equation}
  \underbrace{\frac{1}{16} \cdot \ceil*{\frac{\OupDepth}{\OupDepthStack}} \cdot \InpDepth \cdot \InpWidth\sp{2}}\sb{\text{input depth slices}} + \underbrace{\OupDepth \cdot \InpDepth \cdot \FiltWidth\sp{2}}\sb{\text{filter parameters}}.
\end{equation}
The number of words stored to main memory remains $\OupDepth \cdot \OupWidth\sp{2}$.

\subsubsection{Compute-to-Communication Ratio}

The overall \gls{ccr} is not affected by loading most input depth slices from other clusters instead of main memory.
However, the \gls{ccr} considering only off-chip main memory is significantly increased due to the substantially reduced number of words loaded from main memory.

To give some numerical intuition for a typical layer with $\InpWidth = \OupWidth = 32$, $\FiltWidth = 3$, and $\InpDepth = \OupDepth = 128$.
For single-precision data, \OupDepthStack may be at most 23, and the \gls{ccr} considering only off-chip memory accesses would then be \SI{541.4}{\mac\per\word} or \SI{270.7}{\spflop\per\byte}.
For double-precision data, \OupDepthStack may be at most 11, and the \gls{ccr} considering only off-chip memory accesses would then be \SI{540.6}{\mac\per\word} or \SI{135.2}{\dpflop\per\byte}.

\subsubsection{Summary}

\Cref{alg:conv:par_oup_depth_slices_stack_inp_depth_slice_reuse} improves \cref{alg:conv:par_oup_depth_slices_stack} in that it loads input depth slices that have already been loaded by another cluster in the same L2 quadrant from that cluster instead of off-chip main memory.
This significantly reduces the off-chip memory traffic.

\section{Fully-Connected Layers}

\subsection{Parallelize Input Depth Slices over Clusters}

\begin{algorithm}[htbp]
  \AlgHeaderFc

  initialize a private output volume $\OupVol'$ on each cluster to zero\;
  \ParFor{$d\sb{i} \leftarrow 0$ \KwTo $\InpDepth$}{%
    \DmaLoad{$\InpVol[:,:,d\sb{i},:]$} from main memory\;
    \DmaLoad{$\FiltParams[:,:,d\sb{i},0]$} from main memory\;
    \DmaWait{$\InpVol[:,:,d\sb{i},:]$}\;
    \For{$d\sb{o} \leftarrow 0$ \KwTo $\OupDepth$}{%
      \If{$d\sb{o} + 1 < \OupDepth$}{%
        \DmaLoad{$\FiltParams[:,:,d\sb{i},d\sb{o}+1]$} from main memory\;
      }
      \DmaWait{$\FiltParams[:,:,d\sb{i},d\sb{o}]$}\;
      \For{$b \leftarrow 0$ \KwTo $\BatchSize$}{%
        $\OupVol'[0,0,d\sb{o},b]$ \pluseq \ElemMac{$\InpVol[:,:,d\sb{i},b]$, $\FiltParams[:,:,d\sb{i},d\sb{o}]$}\;
      }
    }
  }
  sum-reduce the private output volumes $\OupVol'$ of all clusters to $\OupVol$\;
  \DmaStore{$\OupVol[:,:,:]$} to main memory\;
  
  \AlgFooter{Implementation of a fully-connected layer where the depth slices of the input volumes are parallelized over the clusters.}%
  \label{alg:fc:par_inp_depth_slices}
\end{algorithm}

The implementation of a fully-connected layer shown in \cref{alg:fc:par_inp_depth_slices} parallelizes the input depth slices over the clusters.
Before the parallel section, each cluster allocates a private output volume $\OupVol'$ and initializes it to zero.
In the parallel section, the cluster first loads the entire batch of one depth slice of the input volume and then loops over the output depth slices.
Within that loop, the cluster loads the filter parameters for the current pair of input and output depth slices and then enters an inner loop over the batch.
Within the inner loop, the cluster in the \ElemMac{} operation multiplies the input depth slice of a batch element with the loaded filter parameters element-wise and then accumulates all products to a single value, which it adds to the output element for the current output depth slice and batch element.
After the parallel section, the private output volumes of all clusters are reduced by summation to a single output volume, which contains the contributions of all input depth slices.

This implementation exploits batching of input and output data, so the filter parameters for each pair of input and output depth slices can be used $\BatchSize$ times.
This is crucial for increasing the operational intensity to a range where the implementation is not memory-bound.

The main problem of this implementation, however, is that each cluster must store a private copy of the entire output volume, which are $\OupDepth \cdot \BatchSize$ words.
This limits the maximum output depth (which restricts the generality of the implementation) or the batch size (which limits the operational intensity).
The implementation in \cref{sec:fc:par_inp_depth_slices_stack_oup_depth_slices} will resolve this limitation by processing output depth slices in stacks.

\subsubsection{Compute Complexity}%
\label{sec:fc:par_inp_depth_slices:compute}
The element-wise \gls{mac} operation (\ElemMac{}) does $\InpWidth \cdot \InpWidth$~\glspl{mac}, and it is placed inside two loops, so it is executed $\OupDepth \cdot \BatchSize$ times per cluster task.
In total, each cluster task does $\InpWidth\sp{2} \cdot \BatchSize \cdot \OupDepth$~\glspl{mac}.
Together, the clusters do $\InpWidth\sp{2} \cdot \BatchSize \cdot \OupDepth \cdot \InpDepth$~\glspl{mac}.

\subsubsection{Space Complexity}%
\label{sec:fc:par_inp_depth_slices:space}
Each cluster must store:
\begin{enumerate}[(1)]
  \item its private output volume, which are $\OupDepth \cdot \BatchSize$ words;
  \item the entire batch of one input depth slice, which are $\InpWidth\sp{2} \cdot \BatchSize$ words;
  \item the filter parameters for one input-output depth slice pair, which are $\InpWidth\sp{2}$ words.
\end{enumerate}
Thus, the minimum local memory required per cluster is $\OupDepth \cdot \BatchSize + \InpWidth\sp{2} \cdot (\BatchSize + 1)$ words.
For a typical layer with $\InpWidth = 7$ and $\OupDepth = 4096$ and a reasonable batch size of $\BatchSize = 32$, this corresponds to \num{132689} words, or ca.\ \SI{519}{\kibi\byte} for single-precision and ca.\ \SI{1037}{\kibi\byte} for double-precision data.
This is too much for a cluster with \SI{128}{\kibi\byte} of local memory.
Decreasing the batch size below 32 usually leads into a memory-bound regime and is therefore not an option.
However, this algorithm can be viable up for $\OupDepth$ on the order of 512 for single-precision data and $\OupDepth$ on the order of 256 for double-precision data; more on that below.

\BufStorage%
  {the filter parameters (3)}%
  {very short inner loop iteration}%
  {input depth slices (2)}
In total, roughly \SI{32}{\kibi\byte} are required as buffers for input depth slices and filter parameters, which leaves \SI{96}{\kibi\byte} for the cluster-private output volume.
For $\InpWidth = 7$ and $\BatchSize = 32$, this allows $\OupDepth \leq 768$ for single-precision and $\OupDepth \leq 384$ for double-precision data.

\subsubsection{Communication Complexity}

\paragraph{Inter-Cluster Communication}%
\label{sec:fc:par_inp_depth_slices:communication}
Before and inside the parallel section, the clusters do not communicate between them.
After the parallel section, the private output volumes $\OupVol'$ of all clusters must to be accumulated to one output volume $\OupVol$.
The optimal way in terms of latency to do this is by tree reduction:
In the first step, for every two clusters in an L1 quadrant, one reads $\OupVol'$ of the other cluster and adds it to its own.
In the second step, one of the two clusters in an L1 quadrant that read $\OupVol'$ in step 1 reads it from the other cluster and adds it to its own.
This then goes on until one cluster in one L3 quadrant reads the accumulated $\OupVol'$ of the other L3 and adds it to its own.
In the first two steps, $\OupVol'$ is read three times over each L1 network.
In steps three and four, $\OupVol'$ is read three times over each L2 network.
In steps five and six, $\OupVol'$ is read three times over each L3 network.
And in the last step, $\OupVol'$ is read once between L3 networks.
In total, $127 \cdot \OupDepth \cdot \BatchSize$ words are communicated between clusters to reduce all $\OupVol'$s into $\OupVol$.

\paragraph{Main Memory Accesses}
Inside the parallel section, each cluster loads the entire batch of one input depth slice, which are $\InpWidth\sp{2} \cdot \BatchSize$ words, from main memory.
Then, in a loop of $\OupDepth$ iterations, each cluster loads the filter parameters for the current pair of input and output depth slices, which are $\InpWidth\sp{2}$ words, from main memory.
Thus, within every cluster task, each cluster loads $\InpWidth\sp{2} \cdot (\BatchSize + \OupDepth)$ words from main memory.
Together, the clusters in all iterations load $\InpDepth \cdot \InpWidth\sp{2} \cdot (\BatchSize + \OupDepth)$ words from main memory.
After the parallel section, one cluster stores the accumulated output volume, $\OupVol$, which are $\OupDepth \cdot \BatchSize$ words, to main memory.

\subsubsection{Compute-to-Communication Ratio}

The bulk of accesses to main memory are the loads by each cluster within the parallel region.
Dividing the compute complexity by that memory access complexity gives the \gls{ccr} within the parallel region:
\begin{equation}
  \text{\acrshort{ccr}} = \frac{\InpWidth\sp{2} \cdot \BatchSize \cdot \OupDepth}{\InpWidth\sp{2} \cdot (\BatchSize + \OupDepth)} \frac{\text{MAC}}{\text{word}} = \frac{\BatchSize \cdot \OupDepth}{\BatchSize + \OupDepth} \frac{\text{MAC}}{\text{word}}.%
  \label{eq:fc:par_inp_depth_slices:ccr}
\end{equation}
As typically $\OupDepth \gg \BatchSize$, the \gls{ccr} tends to scale linearly with the batch size.
\Cref{eq:fc:par_inp_depth_slices:ccr} also describes the overall \gls{ccr}, because both compute and communication complexity are multiplied by the factor $\InpDepth$.

Clearly, the \gls{ccr} increases with increasing batch size $\BatchSize$ and output volume depth $\OupDepth$.
To give some numerical intuition for $\BatchSize = 32$:
with single-precision data, $\OupDepth$ may be at most 768 and the \gls{ccr} would then be \SI{30.7}{\mac\per\word} or \SI{15.4}{\spflop\per\byte};
and with double-precision data, $\OupDepth$ may be at most 384 and the \gls{ccr} would then be \SI{29.5}{\mac\per\word} or \SI{7.4}{\dpflop\per\byte}.

\subsubsection{Summary}

\Cref{alg:fc:par_inp_depth_slices} implements fully-connected layers with a favorable \gls{ccr}, which scales linearly with the batch size, and it provides large parallel sections within which no inter-cluster communication is required.
The main limitation of this algorithm is the local memory required to hold a cluster-private copy of the output volume, which limits it to $\OupDepth \leq 768$ for single-precision and $\OupDepth \leq 384$ for double-precision data.

\subsection{Stacks of Output Depth Slices and Parallel Input Depth Slices}%
\label{sec:fc:par_inp_depth_slices_stack_oup_depth_slices}

\begin{algorithm}[htbp]
  \AlgHeaderFc

  \For{$\delta\sb{o} \leftarrow 0$ \KwTo $\ceil{\OupDepth / \OupDepthStack}$}{%
    $D\sb{O,\mathrm{begin}} \leftarrow \delta\sb{o}\cdot\OupDepthStack$\;
    $D\sb{O,\mathrm{end}} \leftarrow \min(D\sb{O,\mathrm{begin}} + \OupDepthStack, \OupDepth)$\;
    initialize a private output volume $\OupVol'[:,:,D\sb{O,\mathrm{begin}}:D\sb{O,\mathrm{end}},:]$ on each cluster to zero\;
    \ParFor{$d\sb{i} \leftarrow 0$ \KwTo $\InpDepth$}{%
      \DmaLoad{$\InpVol[:,:,d\sb{i},:]$} from main memory\;
      \DmaLoad{$\FiltParams[:,:,d\sb{i},D\sb{O,\mathrm{begin}}]$} from main memory\;
      \DmaWait{$\InpVol[:,:,d\sb{i},:]$}\;
      \For{$d\sb{o} \leftarrow D\sb{O,\mathrm{begin}}$ \KwTo $D\sb{O,\mathrm{end}}$}{%
        \If{$d\sb{o} + 1 < D\sb{O,\mathrm{end}}$}{%
          \DmaLoad{$\FiltParams[:,:,d\sb{i},d\sb{o}+1]$} from main memory\;
        }
        \DmaWait{$\FiltParams[:,:,d\sb{i},d\sb{o}]$}\;
        \For{$b \leftarrow 0$ \KwTo $\BatchSize$}{%
          $\OupVol'[0,0,d\sb{o},b]$ \pluseq \ElemMac{$\InpVol[:,:,d\sb{i},b]$, $\FiltParams[:,:,d\sb{i},d\sb{o}]$}\;
        }
      }
    }
    sum-reduce the private output volumes $\OupVol'[:,:,D\sb{O,\mathrm{begin}}:D\sb{O,\mathrm{end}},:]$ of all clusters to $\OupVol[:,:,D\sb{O,\mathrm{begin}}:D\sb{O,\mathrm{end}},:]$\;
    \DmaStore{$\OupVol[:,:,D\sb{O,\mathrm{begin}}:D\sb{O,\mathrm{end}},:]$} to main memory\;
  }

  \AlgFooter{Implementation of a fully-connected layer where the depth slices of the output volume are processed in stacks of size $\OupDepthStack$ and the depth slices of the input volumes are parallelized over the clusters.}%
  \label{alg:fc:par_inp_depth_slices_stack_oup_depth_slices}
\end{algorithm}

The implementation of a fully-connected layer shown in \cref{alg:fc:par_inp_depth_slices_stack_oup_depth_slices} extends \cref{alg:fc:par_inp_depth_slices} by adding an outer loop that processes the output depth slices in stacks: each cluster processes $\OupDepthStack$ output depth slices in each parallel task.
Within the outer loop, the input depth slices are processed in parallel over the clusters, as in \cref{alg:fc:par_inp_depth_slices}.
This algorithm can be seen as a generalization of \cref{alg:fc:par_inp_depth_slices}: one sets $\OupDepthStack$ below the limit identified in \cref{sec:fc:par_inp_depth_slices:space} and then loops over stacks of output depth slices, executing \cref{alg:fc:par_inp_depth_slices} on each stack of output depth slices.

\subsubsection{Compute Complexity}
The compute complexity is analogous to that in \cref{sec:fc:par_inp_depth_slices:compute}: for each parallel task, each cluster does $\InpWidth\sp{2} \cdot \BatchSize \cdot \OupDepthStack$ \glspl{mac}.
(If $\OupDepthStack$ does not evenly divide $\OupDepth$, the last iteration of the outermost loop processes $\OupDepth \bmod \OupDepthStack$ instead of $\OupDepthStack$ depth slices.)
Over all iterations of the outermost loop, the clusters together do $\InpWidth\sp{2} \cdot \BatchSize \cdot \OupDepth \cdot \InpDepth$ \glspl{mac}.
Thus, in comparison to \cref{alg:fc:par_inp_depth_slices}, this algorithm does not add additional \glspl{mac}.

\subsubsection{Space Complexity}
The space complexity is analogous to that in \cref{sec:fc:par_inp_depth_slices:space}, with $\OupDepth$ replaced by $\OupDepthStack$.
Thus, the minimum local memory required per cluster is $\OupDepthStack \cdot \BatchSize + \InpWidth\sp{2} \cdot (\BatchSize + 1)$ words.
Accounting for buffers for \gls{dma} transfers and setting $\BatchSize = 32$, a typical value of $\InpWidth = 7$ allows for $\OupDepthStack \leq 768$ for single-precision and $\OupDepthStack \leq 384$ for double-precision data.

\subsubsection{Communication Complexity}
\paragraph{Inter-Cluster Communication}
The inter-cluster communication is analogous to that in \cref{sec:fc:par_inp_depth_slices:communication}: in total, $127 \cdot \OupDepth \cdot \BatchSize$ words are communicated between clusters to reduce all $\OupVol'$s into $\OupVol$.

\paragraph{Main Memory Accesses}
The accesses to main memory are similar to those discussed in \cref{sec:fc:par_inp_depth_slices:communication}, with one important difference:
Whereas \cref{alg:fc:par_inp_depth_slices} loaded each input depth slice only once for all output depth slices, \cref{alg:fc:par_inp_depth_slices_stack_oup_depth_slices} loads each depth slice of $\InpVol$ once \emph{per stack of output depth slices}.
This overhead is a result of the trade-off for reduced local memory requirements that we struck by processing the output depth slices in stacks.
Within every cluster task, each cluster loads $\InpWidth\sp{2} \cdot (\BatchSize + \OupDepthStack)$ words from main memory.
Together, the clusters load $\InpDepth \cdot \InpWidth\sp{2} \cdot (\BatchSize + \OupDepthStack)$ words from main memory in every iteration of the outermost loop, and one cluster stores the current depth slice stack of the accumulated output volume, which are $\OupDepthStack \cdot \BatchSize$ words, to main memory.
As there are $\ceil{\OupDepth/\OupDepthStack}$ iterations of the outermost loop, the total number of words loaded from main memory is:
\begin{equation}
  \underbrace{\ceil*{\frac{\OupDepth}{\OupDepthStack}} \cdot \InpDepth \cdot \BatchSize \cdot \InpWidth\sp{2}}\sb{\text{batch of input volumes}} + \underbrace{\OupDepth \cdot \InpDepth \cdot \InpWidth\sp{2}}\sb{\text{filter parameters}}.%
  \label{eq:fc:par_inp_depth_slices_stack_oup_depth_slices:loads_1}
\end{equation}
Thus, the entire input volume is loaded $\ceil{\OupDepth / \OupDepthStack}$ times instead of once as in \cref{alg:fc:par_inp_depth_slices}, but each filter parameter is still loaded only once.
For the same reason, the total number of words stored to main memory remains $\OupDepth \cdot \BatchSize$.
For the sake of computing the \gls{ccr}, we rewrite \cref{eq:fc:par_inp_depth_slices_stack_oup_depth_slices:loads_1} as
\begin{equation}
  \InpDepth \cdot \InpWidth\sp{2} \cdot \left( \ceil*{\frac{\OupDepth}{\OupDepthStack}} \cdot \BatchSize + \OupDepth \right).%
  \label{eq:fc:par_inp_depth_slices_stack_oup_depth_slices:loads_2}
\end{equation}

\subsubsection{Compute-to-Communication Ratio}
The bulk of accesses to main memory are again the loads of the input volume and the filter parameters; the total number of loaded words is given by \cref{eq:fc:par_inp_depth_slices_stack_oup_depth_slices:loads_2}.
Dividing the compute complexity by those loads gives the \gls{ccr}:
\begin{equation}
  \text{\acrshort{ccr}} = \frac{\InpDepth \cdot \InpWidth\sp{2} \cdot \BatchSize \cdot \OupDepth }{\InpDepth \cdot \InpWidth\sp{2} \cdot \left( \ceil*{\frac{\OupDepth}{\OupDepthStack}} \cdot \BatchSize + \OupDepth \right)} \frac{\text{MAC}}{\text{word}} = \frac{\BatchSize \cdot \OupDepth}{\ceil*{\frac{\OupDepth}{\OupDepthStack}} \cdot \BatchSize + \OupDepth} \frac{\text{MAC}}{\text{word}}.
\end{equation}
The overhead of the redundant input volume loads also manifests itself in the denominator (i.e., in the communication) of the \gls{ccr}.

To give some numerical intuition for $\BatchSize = 32$ and $\OupDepth = 4096$:
with single-precision data, $\OupDepthStack$ may be at most 768 and the \gls{ccr} would then be \SI{30.6}{\mac\per\word} or \SI{15.3}{\spflop\per\byte};
and with double-precision data, $\OupDepthStack$ may be at most 384 and the \gls{ccr} would be \SI{29.5}{\mac\per\word} or \SI{7.4}{\dpflop\per\byte}.
Clearly, the impact on the \gls{ccr} of the overhead of processing output depth slices in stacks diminishes for large $\OupDepth$.

\subsubsection{Summary}
\Cref{alg:fc:par_inp_depth_slices_stack_oup_depth_slices} is a generalization of \cref{alg:fc:par_inp_depth_slices} in that the depth of the output volume is no longer limited by the local memory of each cluster.
This algorithm maintains the large parallel sections as well as the favorable \gls{ccr} of \cref{alg:fc:par_inp_depth_slices} for large output volume depths.

\section{Summary}

This document has presented three implementations of convolutional layers and two implementations of fully-connected layers on the Manticore cluster-based many-core architecture.
By cleverly using the local memory of each cluster and the high-performance on-chip network, two convolutional layer implementations and one fully-connected layer implementation attain a high \gls{ccr}, which allow them to exploit the full computational potential of Manticore.

\bibliographystyle{IEEEtran}
\bibliography{main} 

\end{document}